\documentclass[onecollarge,natbib]{svjour2}

\bibpunct{[}{]}{;}{n}{}{,} 
\smartqed  
\usepackage{graphicx}
\usepackage{amsmath,stackrel}

\journalname{Few-Body Systems}
\newcommand{\be}{\begin{equation}}
\newcommand{\ee}{\end{equation}}
\newcommand{\beq}{\begin{eqnarray}}
\newcommand{\eeq}{\end{eqnarray}}

\newcommand{\Dlr}{\buildrel \leftrightarrow \over D\raise-1pt\hbox{}}

\begin{document}

\title{Parton distribution functions from Lattice QCD
}


\author{Constantia Alexandrou      
}


\institute{Constantia Alexandrou \at
           Department Physics, University of Cyprus, P.O. Box 20537, 1678 Nicosia, Cyprus, and  \\
Computation-based Science and Technology Research Research, The Cyprus Institute,20 Kavafi Str.,
  Nicosia 2121,
  Cyprus \\
              Tel.: +357-22-892829\\
               \email{alexand@ucy.ac.cy}           
}

\date{}

\maketitle

\begin{abstract}
We present recent results on the first moments of parton distributions using gauge configurations generated with two degenerate flavors of light twisted mass quarks with pion mass fixed approximately to its physical value. We also present a first study of the vector parton distribution function using a twisted mass
 ensemble  at pion mass of 373~MeV.  
\end{abstract}

\section{Introduction}
\label{intro}
\begin{figure}
\begin{minipage}{0.33\linewidth}
  \includegraphics[width=\linewidth]{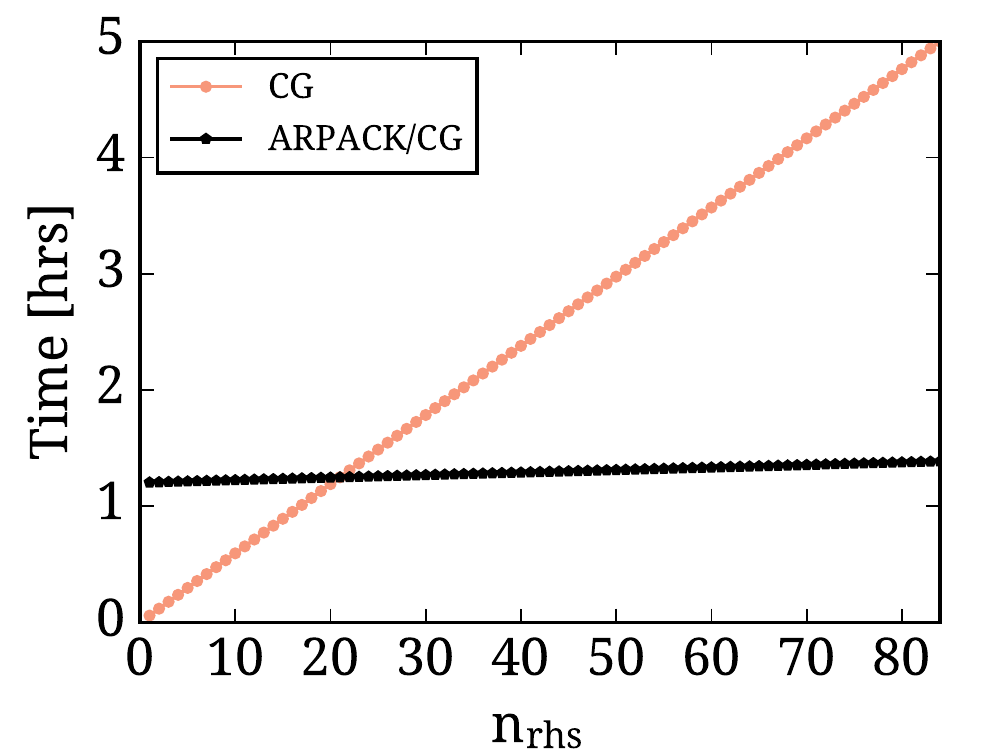}
\end{minipage}\hfill
\begin{minipage}{0.33\linewidth}
\includegraphics[width=\linewidth]{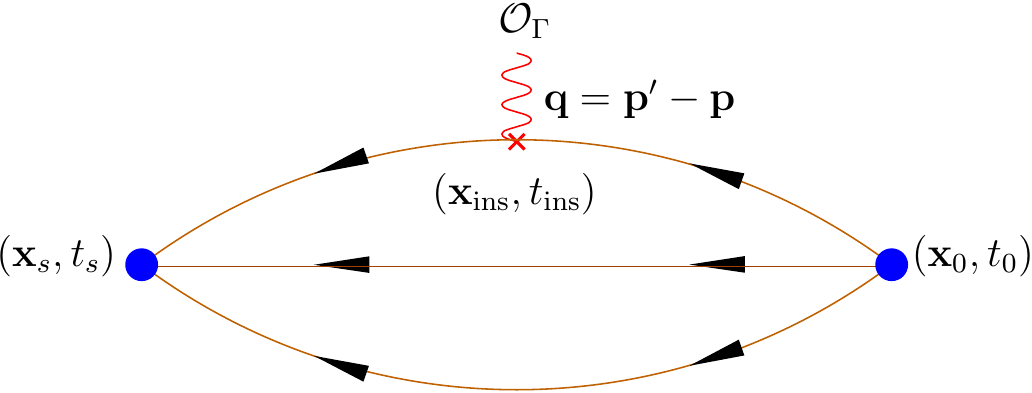}
\end{minipage}\hfill
\begin{minipage}{0.33\linewidth}
\includegraphics[width=\linewidth]{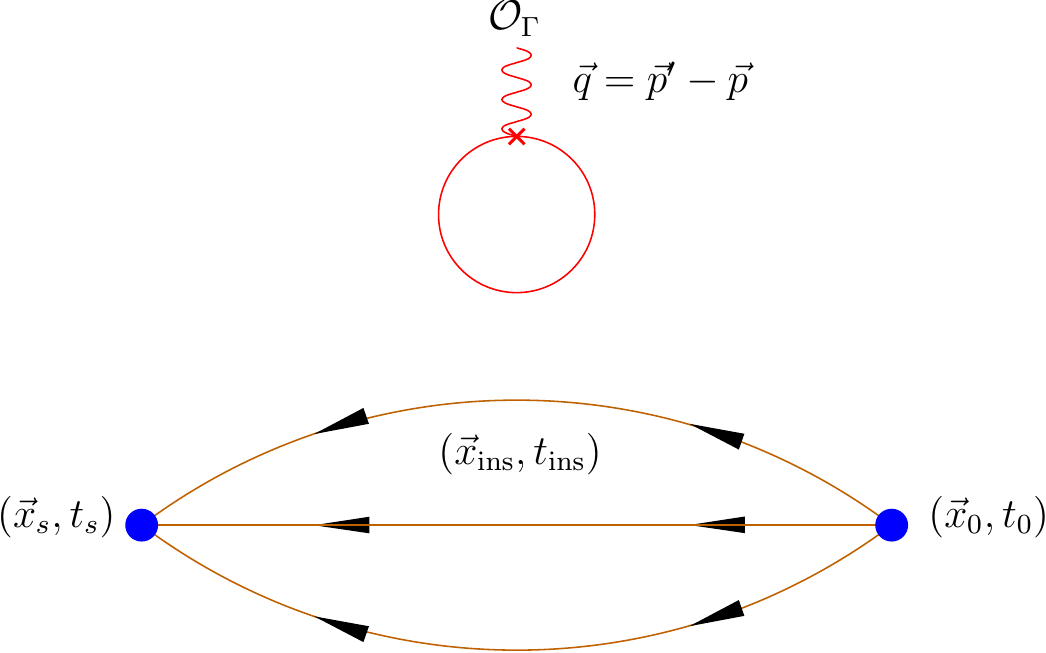}
\end{minipage}
\label{fig:diagrams}       
\caption{Speedup achieved with exact deflation as a function of the number of rhs $n_{\rm rhs}$(left). Connected (middle) and disconnected (right) contributions to a three-point function.}
\end{figure}

Although  nucleon structure has been an area of intensive activity
within lattice QCD for a number of years, it is only recently that simulations   with near physical parameters both in
terms of the value of the pion mass (referred to as physical point), as well as, with respect to the lattice volume and lattice spacing
 have become available~\cite{Abdel-Rehim:2015owa,Abdel-Rehim:2015pwa,Bali:2016wqg,Bhattacharya:2015wna,Durr:2015dna,Ohta:2014rfa,Green:2014xba,Horsley:2013ayv,Green:2012ud}.  Generalized parton
distributions (GPDs) encode information related to nucleon structure
that complements the information extracted from form
factors~\cite{Mueller:1998fv,Ji:1996nm,Radyushkin:1997ki}.  Their forward limit
coincides with the usual parton distribution functions (PDFs) and, using Ji's sum
rule~\cite{Ji:1996ek}, allows one to determine the contribution of a
specific parton to the nucleon spin. In the context of the
``proton spin puzzle'', which refers to the unexpectedly small fraction of
the total spin of the nucleon carried by quarks, this has triggered
intense experimental activity~\cite{Airapetian:2009rj, Chekanov:2008vy, Aaron:2007cz,
MunozCamacho:2006hx, Stepanyan:2001sm}.
Calculations using gauge configurations produced with  physical values of the light quark mass, pose new challenges  associated with the slower convergence of the inversion of the Dirac operator and the increase of gauge noise requiring very large  computational resources. Therefore, techniques to speedup the inversions are critical in order to obtain results at the physical point. In Fig.~\ref{fig:diagrams} we show the speedup achieved when deflation of lowest eigen-modes is applied. The best improvement is achieved with exact deflation of the lowest about 1800 eigen-modes, which yields a speed-up of a factor of 20 as compared to the conjugate gradient method after about 1000 right hand sides (rhs)  that is the typical number of rhs in our production runs.

\section{Evaluation of matrix elements in lattice QCD}
\label{sec:1}
In Fig.~\ref{fig:diagrams} we show the contributions entering the evaluation of a hadron matrix element. One needs to compute the  three-point function given by

\be G^{\mu\nu}({\Gamma},\vec q,t_s, t_{\rm ins}) =\sum_{\vec x_s, {\vec x}_{\rm ins}} \, e^{i{\vec x}_{\rm ins} \cdot \vec q}\, 
     {\Gamma_{\beta\alpha}}\, \langle {J_{\alpha}(\vec x_s,t_s)} {\cal O}_\Gamma^{\mu\nu}({\vec x}_{\rm ins},t_{\rm ins}) {\overline{J}_{\beta}(\vec{x}_0, t_0)} \rangle 
\ee
and then form a ratio by dividing the three-point function by an appropriate combination of two-point functions:
 \be
    R(t_s,t_{\rm ins},t_0) \begin{array}{c}  {\scriptstyle (t_s-t_{\rm ins})\Delta \gg 1,(t_{\rm ins}-t_0)\Delta \gg 1}\\{\longrightarrow {\longrightarrow} \longrightarrow} \\{}\end{array} \mathcal{M}\left [1
      + \dots e^{-\Delta({\bf p})(t_{\rm ins}-t_0)} + \dots e^{-\Delta({\bf
          p'})(t_s-t_{\rm ins})}\right].
\label{ratio} 
 \ee
In Eq.~\ref{ratio} $\mathcal{M}$ is the desired matrix element, $t_s,t_{\rm ins},t_0$ the
  sink, insertion and source time-slices and  $\Delta({\bf p})$ the
  energy gap with the first excited state. Summing over the current insertion time $t_{\rm ins}$ we obtain
\be
  \sum_{t_{\rm ins}=t_0}^{t_s} R(t_s,t_{\rm ins},t_0) = {\sf Const.} + \mathcal{M}[(t_s-t_0) + \mathcal{O}(e^{-\Delta({\bf p})(t_s-t_0)})  + \mathcal{O}(e^{-\Delta({\bf p'})(t_s-t_0)})].
\label{summed ratio}
 \ee
In the summed ratio, excited state contributions are suppressed by 
exponentials decaying with $t_s-t_0$, rather than $t_s-t_{\rm ins}$ and/or $t_{\rm ins}-t_0$. However, one needs to fit the slope rather than to a constant, which increases the statistical uncertainty in the extracted value of the desired matrix element. 

In order to  extract the physical value of the matrix element from the lattice results one needs to carry out the renormalization of the lattice matrix element by evaluating the renormalization function $Z(\mu,a)$. The renormalization functions are  typically computed non-perturbatively. In this work we performed a 
perturbative subtraction of ${\cal O}(a^2)$ lattice artifacts, which in general leads to a better determination of the renormalization functions~\cite{Alexandrou:2012mt}.

\section{Generalized Parton Distributions }

The standard procedure to study generalized parton distributions in lattice QCD
is to use 
factorization of the light cone matrix element in terms of local operators. There are three types of GPDs involving the following operators:  i) the vector operator 
${\cal O}_{V^a}^{\mu_1 \cdots \mu_n}=\bar{\psi}(x)\gamma^{\{\mu_1}i\Dlr^{\mu_2}\ldots i\Dlr^{\mu_n \}}\frac{\tau^a}{2}\psi(x)$, 
ii) the axial-vector operator
${\cal O}_{A^a}^{\mu_1 \cdots \mu_n}=\bar{\psi}(x)\gamma^{\{\mu_1}i\Dlr^{\mu_2}\ldots i\Dlr^{\mu_n \}}\gamma_5\frac{\tau^a}{2}\psi(x)$
and iii) the
tensor operator
${\cal O}_{T^a}^{\mu_1 \cdots \mu_n}=\bar{\psi}(x)\sigma^{\{\mu_1,\mu_2}i\Dlr^{\mu_3}\ldots i\Dlr^{\mu_n \}}\frac{\tau^a}{2}\psi(x)$. In the special case when no derivatives are present then one obtains the  nucleon form factors, while for $Q^2=0$ one has the parton distribution functions (PDFs). In this work, we limit 
ourselves to
 one-derivative operators determining the first moments of these PDFs.

Generalized form factor decomposition of the vector one-derivative operator
yields
\be 
\langle N(p^\prime,s^\prime) | {\cal O}_{V^3}^{\mu\nu} {| N(p,s) \rangle} = 
    \bar u_N(p^\prime, s^\prime) 
     \Biggl[  {A_{20}(q^2)} \gamma^{\{\mu}P^{\nu\}}+{B_{20}(q^2)} \frac{i\sigma^{\{\mu \alpha}q_\alpha P^{\nu\}}}{2m}
+C_{20}(q^2) \frac{q^{\{\mu}q^{\nu\}}}{m} \biggr] \frac{1}{2}u_N(p,s) 
\label{vector}
\ee
The
nucleon spin due to a quark $q$ is then determined by $J^{q}=\frac{1}{2}\biggl [A^{q}_{20}(0)+{B^q_{20}(0)}\biggr]$, while the momentum fraction is given by $\langle x \rangle_q=A^q_{20}(0)$.

\subsection{Momentum fraction of the pion and the nucleon}
\begin{figure*}
\begin{minipage}{0.49\linewidth}
\includegraphics[width=\linewidth]{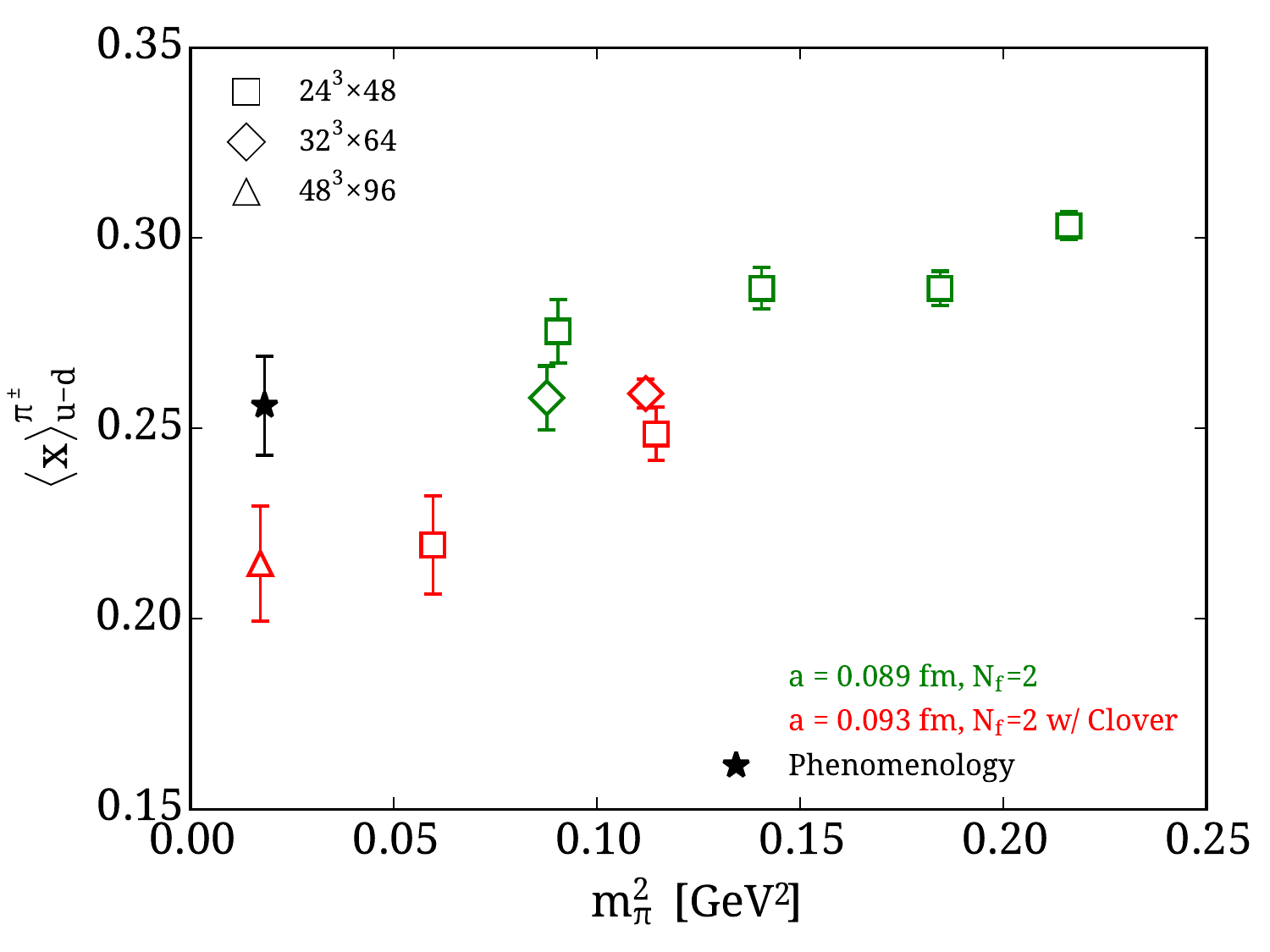}
\end{minipage}\hfill
\begin{minipage}{0.49\linewidth}
\includegraphics[width=\linewidth]{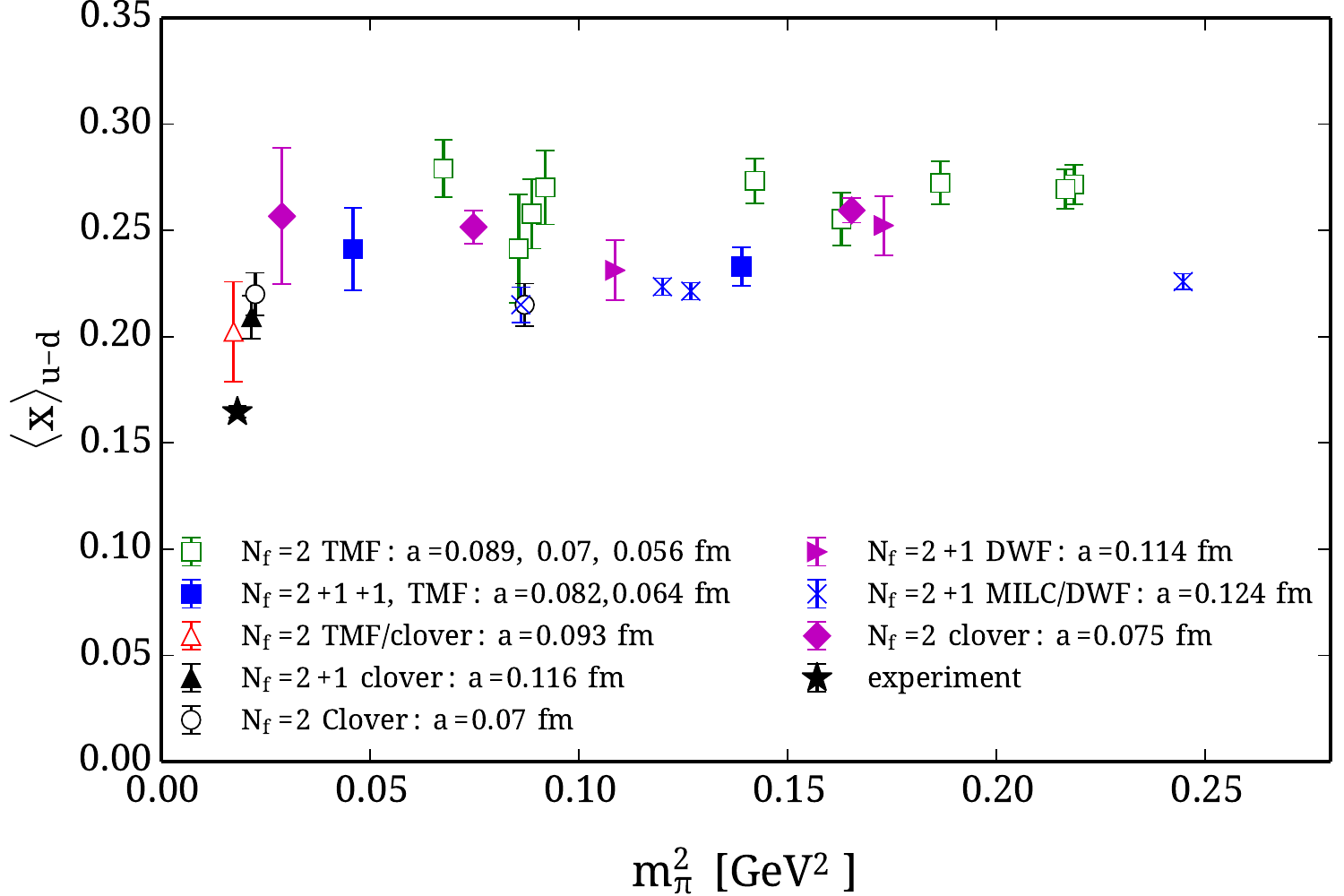}
\end{minipage}
\caption{$\langle x\rangle_{u-d}$ in the $\overline{MS}$ scheme at  $\mu= 2$~GeV for the pion (left) and for the nucleon (right). The experimental value for  $\langle x \rangle_{u-d}$ shown here is from  Ref.~\cite{Alekhin:2012ig}.
}
\label{fig:x}
\end{figure*}

In Fig.~\ref{fig:x}, we show results on
the isovector $\langle x\rangle_{u-d}$  for which disconnected contributions are zero in the isospin limit for the pion and the nucleon in the $\overline{MS}$ scheme at  $\mu= 2$~GeV.

In the case of the pion we show results obtained using $N_f=2$ twisted mass fermions. There is agreement between
the clover-improved and non-clover improved ensembles~\cite{Abdel-Rehim:2015owa} while no volume effects are observed  within our statistical errors.
The phenomenological value is  extracted from a next-to-leading order analysis from Fermilab E-615 pionic Drell-Yan data~\cite{Wijesooriya:2005ir}.
The  nucleon momentum fraction has been studied by a number of collaborations.
In particular, 
near the physical point, in addition to 
 $N_f=2$ twisted mass plus clover-improved   fermions from ETMC~\cite{Abdel-Rehim:2015owa,Abdel-Rehim:2014nka,Abdel-Rehim:2013yaa,Abdel-Rehim:2013wba,Abdel-Rehim:2013yaa}, there are results using $N_f = 2+1$ clover fermions with 2-HEX smearing from LHPC~\cite{Green:2012ud} and   $N_f=2$ clover fermions~\cite{Bali:2014gha,Pleiter:2011gw} from QCDSF/UKQCD.
As can be seen for the latest lattice results close to the physical pion mass,  $\langle x\rangle_{u-d}$  approaches the physical value. Furthermore,
these studies show that for bigger sink-source time  separations the value decreases towards to the physical one. We note that, as the sink-source time separation increases, one needs increasingly larger  statistics  to attain  statistical errors that can yield a meaningful result and this analysis is on-going.

\subsection{Nucleon gluon moment}
We consider the nucleon matrix element of the gluon operator $O_{\mu\nu}=-{\rm Tr}[ G_{\mu\rho} G_{\nu\rho}]$,
where we take  $\langle N|O_{44}-\frac{1}{3}O_{jj}|N\rangle$ at zero momentum, which yields directly $\langle x \rangle_g$. We use HYP-smearing to reduce noise
and, for this operator,  apply  perturbative renormalization.
We analyze two ensembles :
one $N_f=2+1+1$ twisted mass ensemble with $a$ = 0.082~fm, $m_\pi$ = 373~MeV,  using  34,470 statistics and one  $N_f=2$ twisted mass plus clover ensemble, with  $a=0.093$~fm, $m_\pi=132$~MeV and $\sim$155,800 statistics. 
For the latter ensemble, we find $\langle x\rangle_g=0.282(39)$ in ${\overline{\rm MS}}$ scheme at $\mu=2$~GeV, which gives for the first time a determination  of this important quantity 
at the physical value of the  pion mass.

\subsection{Nucleon spin}

Having computed the generalized form factors $A_{20}$ and $B_{20}$ we can 
determine 
the nucleon spin carried by the quarks via the relation $J^q=\frac{1}{2}\left(A^q_{20}+B^q_{20}\right)$. The spin sum  given by $ \frac{1}{2}=\sum_{q}J^q={\left(\frac{1}{2}\Delta \Sigma^q +L^q\right)} +J^G $ includes  a gluon contribution $J^G$. The intrinsic quark spin is given by
$\Delta \Sigma^q=g^q_A$, which has been computed within lattice QCD. Knowing
$J^q$ and $g^q_A$ one can extract the angular momentum $L^q$.

\begin{figure*}
\begin{minipage}{0.49\linewidth}
{\includegraphics[width=\linewidth]{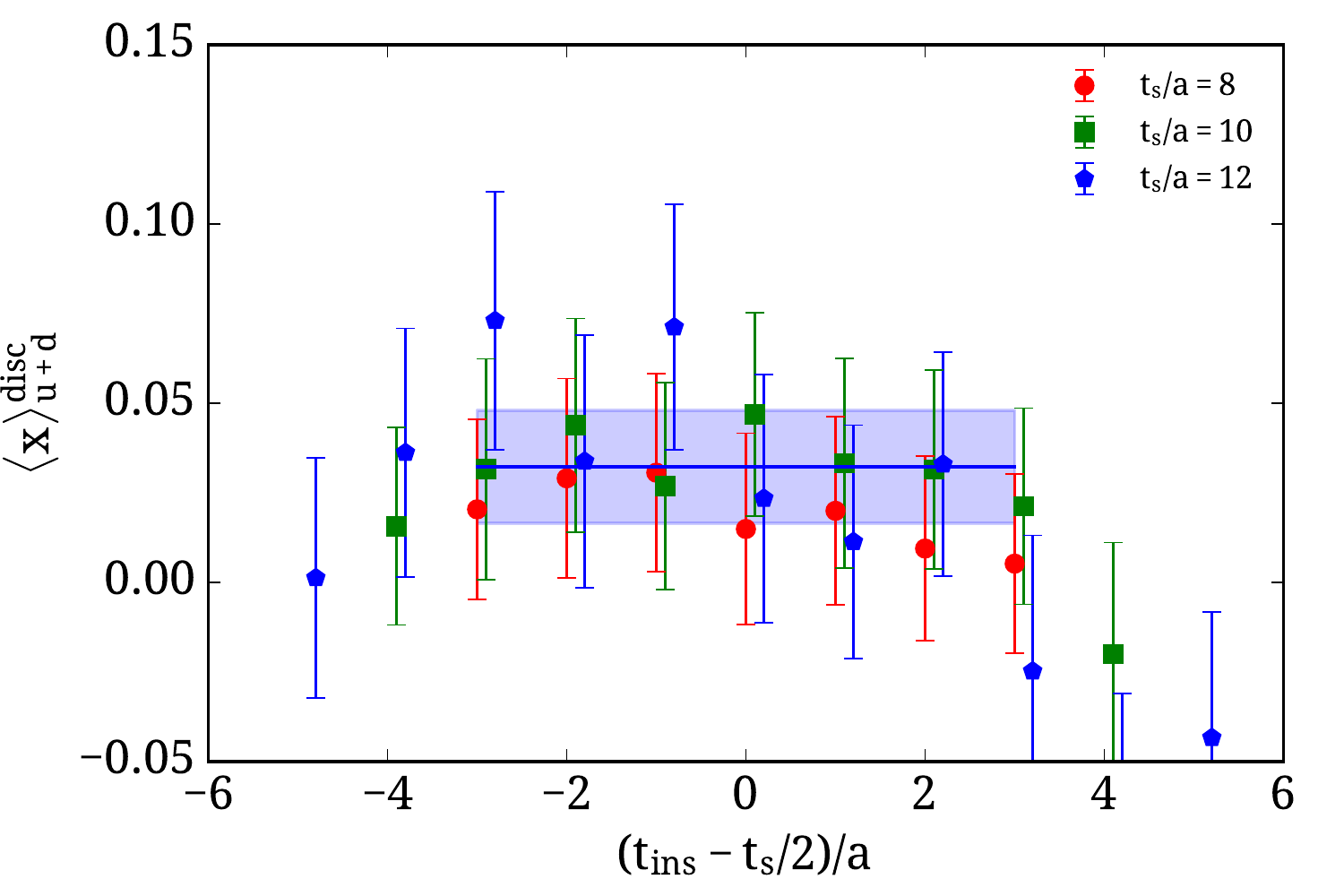}}
\end{minipage}\hfill
\begin{minipage}{0.49\linewidth}
{\includegraphics[width=\linewidth]{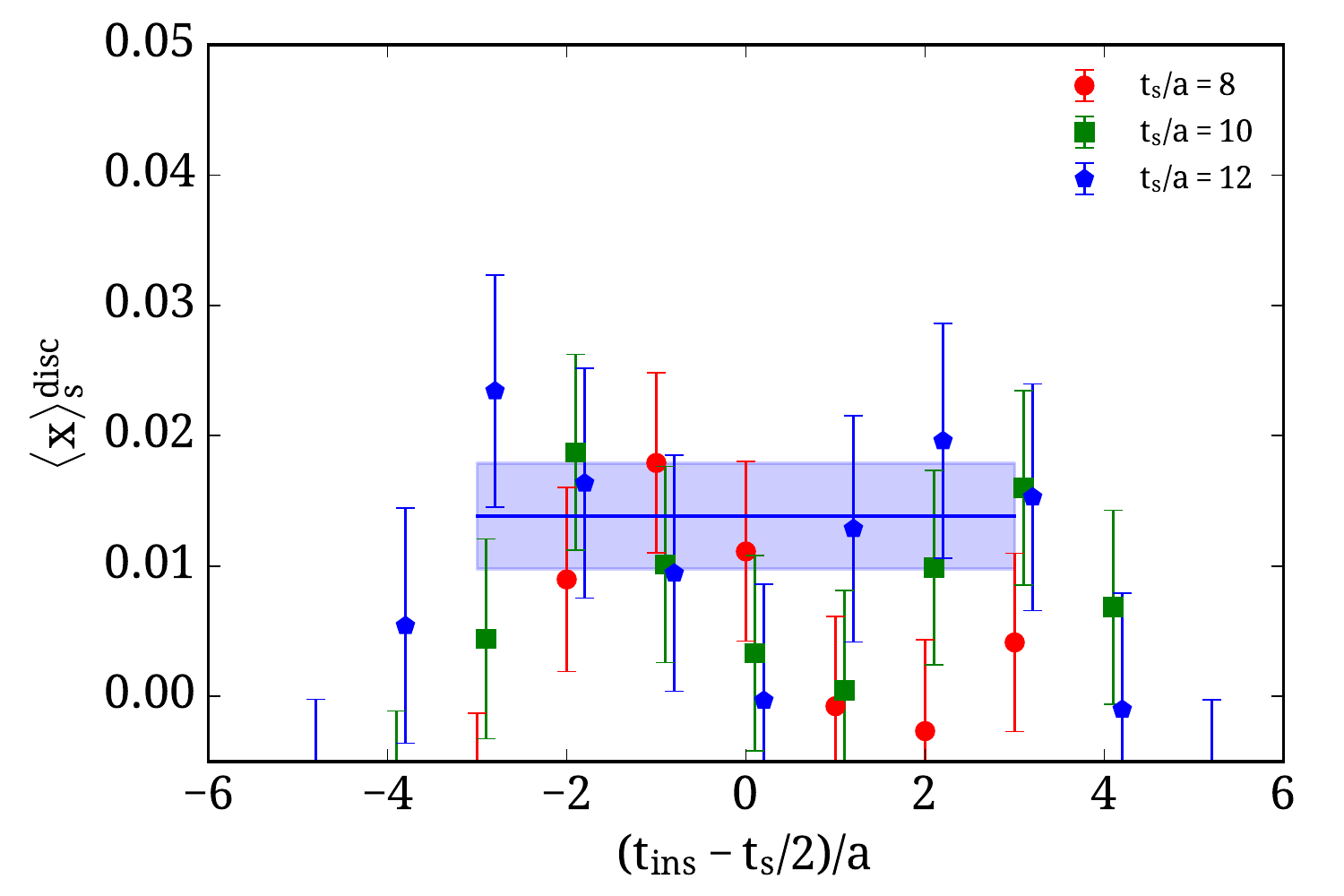}}
\end{minipage}
\caption{Disconnected contributions to the nucleon momentum fraction at the physical point for the light (left) using currently 39000 measurements and for the strange (right) using about 185000 measurements. }
\label{fig:ave x}
\end{figure*}
\begin{figure}[h!]
\begin{minipage}{0.49\linewidth}
{\includegraphics[width=\linewidth]{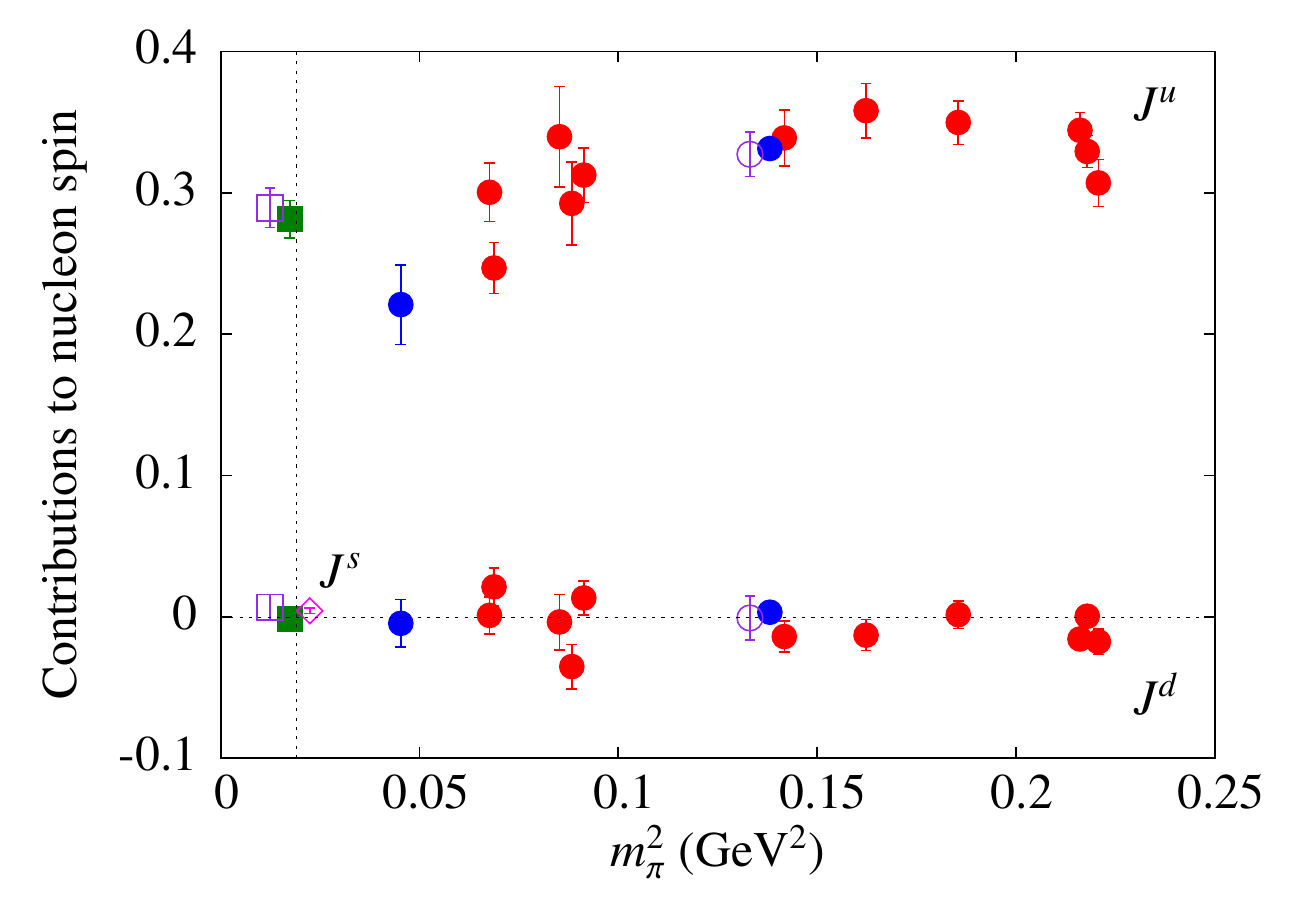}}
\end{minipage}
\begin{minipage}{0.49\linewidth}
{\includegraphics[width=\linewidth]{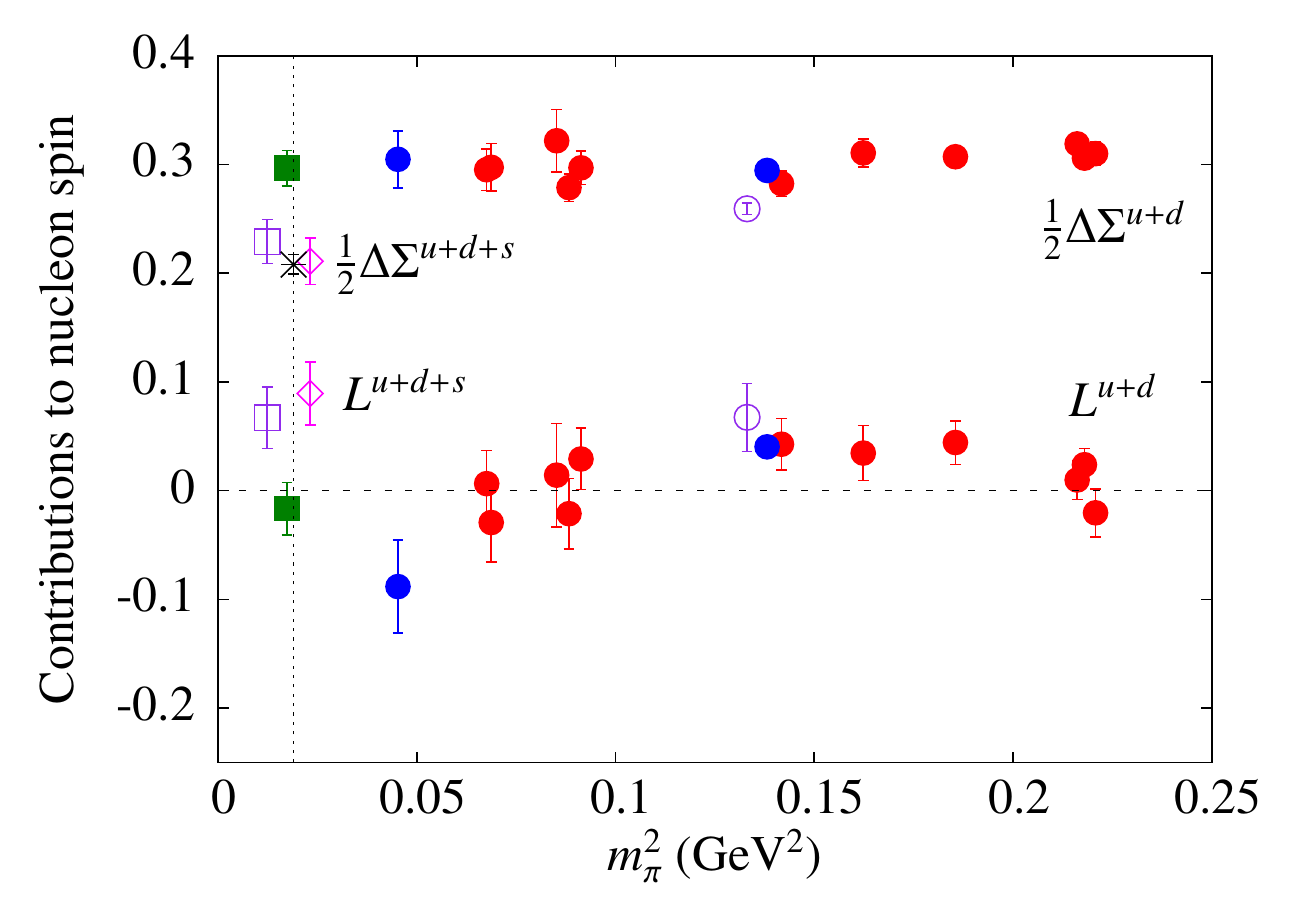}}
\end{minipage}
\caption{The total quark spin (left) carried by the light and the strange quarks and $\Delta \Sigma^{u+d}$  and $L^{u+d}$ (right) in the $\overline{\rm MS}$ scheme at 2~GeV using $N_f=2$ and $N_f=2+1+1$ twisted mass fermions. Open circles and open squares  include disconnected contributions from the u and d quarks, while the open diamonds also include the strange quark contribution. The asterisk is the experimental value of the intrinsic spin.}
\label{fig:spin}
\vspace*{-0.5cm}
\end{figure}

In order to extract the individual quark contribution one needs to compute both
the connected and disconnected parts shown schematically in Fig.~\ref{fig:diagrams}. Connected contributions have been computed using the well established sequential inversion through the sink method.
Disconnected contributions have been computed with new methods developed for graphics cards (GPUs). This was carried out for ensembles using  ${\cal O} (150, 000)$ statistics for the $N_f=2+1+1$ ensemble of twisted mass fermions at $m_\pi=373$~MeV~\cite{Abdel-Rehim:2013wlz} as well as  using an $N_f=2$ ensemble of twisted mass fermions with a clover term at  $m_\pi=132$~MeV~\cite{Abdel-Rehim:2015lha}, which is still  on-going.
We show in Fig.~\ref{fig:ave x} the momentum fraction $\langle x\rangle_q=A^q_{20}(0)$ arising from the disconnected part. As can be seen, it is non-zero for both the light and the strange quarks. Whereas $A_{20}^q(0)$ is accessible directly at $Q^2=0$, $B_{20}(0)$ is not, requiring extrapolation to $Q^2=0$. Although we have computed the disconnected contributions to $B_{20}$ for the few lowest $Q^2$-values the results are consistent with zero and  we will ignore them in what follows. The disconnected intrinsic spin  $\Delta \Sigma^q $ is also found to be non-zero. We show in Fig.~\ref{fig:spin} the total spin $J^q$ as well as $\Delta \Sigma^{u+d}$  at the physical point as well as 
 $\Delta \Sigma^{u+d+s}$. As can be seen, including the disconnected contributions brings agreement with the experimental value in the case of $\Delta\Sigma$. Using $L^q=J^q-\frac{1}{2}\Delta\Sigma^q$ we obtain results on the angular momentum. The major outcome is that disconnected contributions produce a non-zero angular momentum at the physical point.
 At the physical point, our preliminary  values are  $J^{u+d}= 0.296(17)$ and $L^{u+d}=0.067(28)$, while  $\frac{1}{2}\Delta \Sigma^{u+d}=0.229(20)$ and $\frac{1}{2}\Delta \Sigma^{u+d+s}=0.211(21)$, where  for the first time, disconnected contributions are included.

\subsection{Direct evaluation of parton distribution functions - an exploratory study}
\begin{figure}[h!]
\begin{minipage}{0.45\linewidth}
{\includegraphics[width=\linewidth,height=0.9\linewidth]{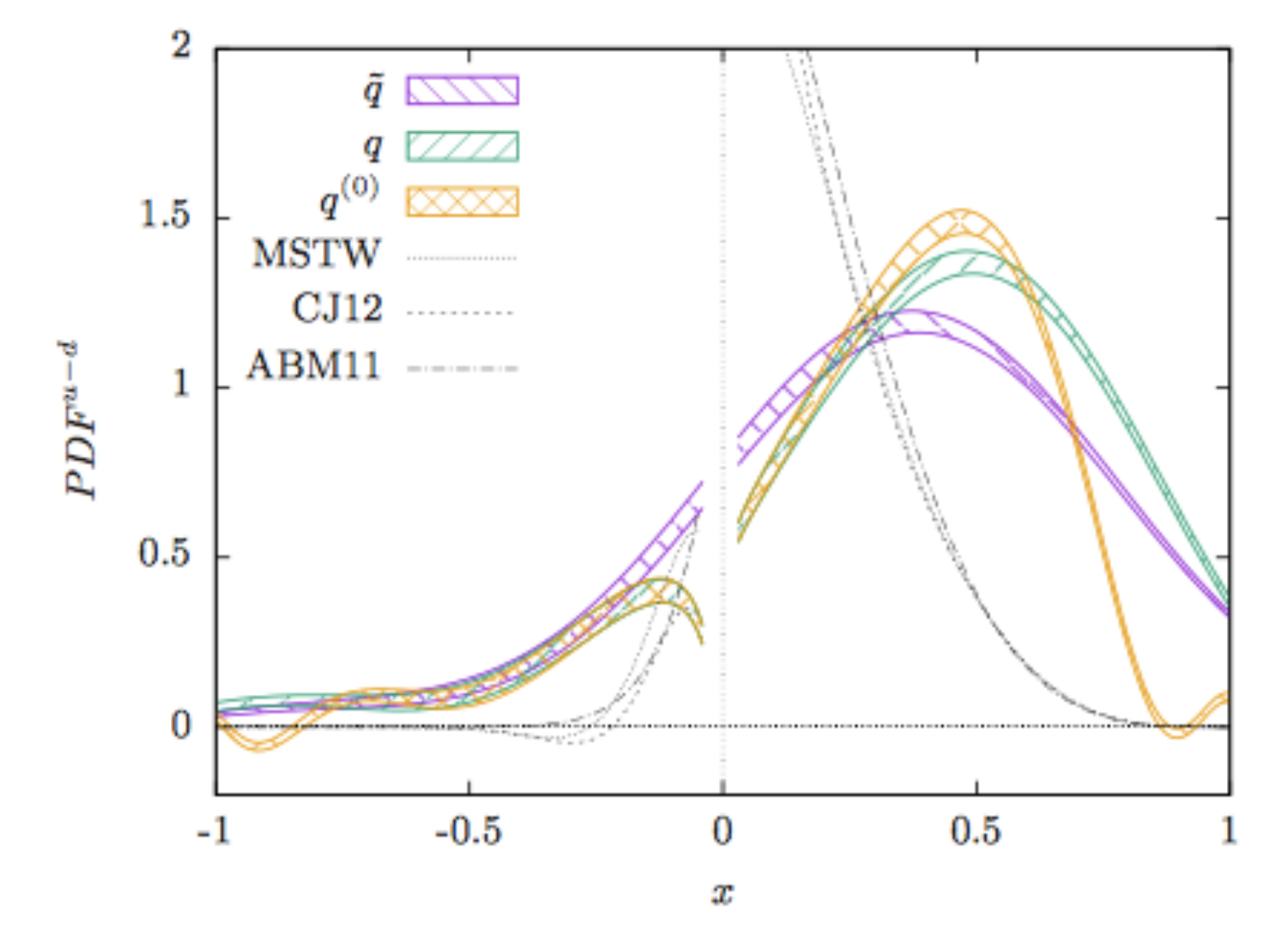}}
\caption{Results on the unrenormalized $q(x)$ for 5-HYP steps, $P_3=4\pi/L$ from Ref.~\cite{Alexandrou:2015rja}.}
\label{fig:PDF}
\end{minipage}\hfill
\begin{minipage}{0.54\linewidth}\vspace*{-1.3cm}
We note that: i)
The calculation of the leading UV divergences in $\tilde{q}$ perturbatively   is done keeping $P_3$ fixed while taking $\Lambda \to \infty$ (in contrast to first taking  $P_3\to\infty$ for the renormalization of $q$); ii)  The renormalization procedure is still under study and thus here 
we  identify the UV regulator as $\mu$ for $q(x)$
and as  $\Lambda$ for the case of the quasi-distribution $\tilde{q}(x)$. The dependence on the UV regulator $\Lambda$ will be translated, in the end, into a renormalization scale $\mu$ after proper renormalization; 
iii) Single pole terms cancel when combining the vertex and wave function corrections, and double poles are  reduced to a single pole that are taken care via the  principal value prescription;
iv) A divergent term remains in $\delta Z^{(1)}$ that depends on the cut-off $x_c$
\end{minipage}
\end{figure}

We consider the matrix element: 
$\tilde{q}(x,\Lambda,P_3)=\int_{-\infty}^{+\infty}  \frac{dz}{4\pi} e^{-izxP_3}{\langle P|\bar{\psi}(z,0)\>\gamma_3 \,W(z)\psi(0,0)|P\rangle}_{h(P_3,z)} $ where $\tilde{q}(x)$
is the quasi-distribution defined in Ref.~\cite{Ji:2013dva}, which
 can be computed in lattice QCD. First results are obtained for $N_f=2+1+1$ clover fermions on HISQ sea~\cite{Lin:2014zya} and for an $N_f=2+1+1$ TMF ensemble with $m_\pi=373~$MeV~\cite{Alexandrou:2015rja} for which we show results in  Fig.~\ref{fig:PDF} on the isovector distribution 
$q^{u-d}(x)$ for 5 steps of HYP smearing. 
The matching to the PDF $q(x)$ is done using 
\be
q(x,\mu)=\tilde{q}(x,\Lambda,P_3)-\frac{\alpha_s}{2\pi}\tilde{q}(x, \Lambda,P_3)\delta Z_F^{(1)}\left(\frac{\mu}{P_3},\frac{\Lambda}{P_3}\right)-\frac{\alpha_s}{2\pi}\int_{-1}^1  \frac{dy}{y} Z^{(1)}\left(\frac{x}{y},\frac{\mu}{P_3},\frac{\Lambda}{P_3}\right)\tilde{q}(y,\Lambda,P_3) +{\cal O}(\alpha_s^2)
\label{PDF}
\ee

\vspace*{-0.5cm}

\section {Conclusions}
 Simulations at near physical parameters of QCD are yielding important results on the structure of hadrons. In this work we have shown results on nucleon observables taking into account for the first time disconnected contributions at the physical point, which are shown to be crucial to obtain agreement with experiment for the intrinsic spin of the nucleon.  Exploration of new techniques to  compute hadron PDFs, charge radii and electric dipole moments is on-going,  as well as, 
the development of techniques for resonances 
and for {\it ab Initio} Nuclear Physics~\cite{Savage:2015eya}. This thus represents a very rich program for zero-temperature hadron and nuclear physics and we expect rapid progress in many of these areas in the near future.

\begin{acknowledgements}
I would like to thank all members of the ETM Collaboration for a most enjoyable collaboration and in particular  A. Abdel-Rehim,  M. Constantinou, K. Jansen, K. Hadjiyiannakou, Ch. Kallidonis, G. Koutsou, B. Kostrzewa, F. Steffens,  C. Urbach, C. Wiese and A. Vaquero for their invaluable contributions to the the results presented here.
This work was supported by a grant from the Swiss National Supercomputing Centre (CSCS) under project ID s540 and in addition used computational resources  from
the John von Neumann-Institute for Computing on the Juropa system and
the BlueGene/Q system Juqueen  at the research center in J\"ulich, resources from the Gauss Centre for Supercopmuting on HazelHen (HLRS) and PRACE project access to the Tier-0 computing resources Curie (CEA), Fermi (CINECA) and SuperMUC (LRZ). 
\end{acknowledgements}


\bibliographystyle{unsrt}
\bibliography{refs}   

%
%

\end{document}